\documentclass{article}

\usepackage{arxiv}

\usepackage[utf8]{inputenc}
\usepackage[T1]{fontenc}
\usepackage{hyperref}
\usepackage{url}
\usepackage{booktabs}
\usepackage{tabularx}
\usepackage{amsfonts}
\usepackage{amsmath}
\usepackage{amsthm}
\newtheorem{proposition}{Observation}
\usepackage{microtype}
\usepackage{courier}
\usepackage[font=small,skip=6pt]{caption}
\usepackage{graphicx}
\usepackage{xcolor}
\hypersetup{colorlinks=true, linkcolor=black, citecolor=blue!50!black, urlcolor=blue!50!black,
  pdftitle={TradeGauntlet: Evaluating Compliance Controls for Autonomous Trading Agents},
  pdfauthor={Sandip Nandi, Aishwarya Bose, Apurba Bose, Soovadeep Bakshi}}

\title{TradeGauntlet: Evaluating Compliance Controls\\for Autonomous Trading Agents}
\date{}
\author{Sandip Nandi\\
  Independent Researcher \And
  Aishwarya Bose\\
  Independent Researcher \AND
  Apurba Bose\\
  Independent Researcher \And
  Soovadeep Bakshi\\
  Independent Researcher}
\renewcommand{\shorttitle}{TradeGauntlet}
\renewcommand{\headeright}{Preprint}
\renewcommand{\undertitle}{Preprint}

\begin{document}
\maketitle

\begin{abstract}
Autonomous trading agents need controls that prevent problematic orders without making legitimate tasks impractical. We study this aspect of agentic compliance with TradeGauntlet, a replayable limit-order-book simulator that separates detection, prevention, and execution cost. In 1,680 simulated sessions with Claude Haiku 5.5 and Sonnet 5.5, all conduct alerts were self-matches. Gates eliminated those alerts by rejecting self-crossing orders, while the aggregate-gate condition reduced Sonnet's two-sided fills by about 15\%. A conduct reminder reduced Sonnet's alert rate but reduced Haiku's fills without a detectable alert benefit. These results show why zero alerts alone do not establish an agent's compliance. Scripted probes and two search methods test weaknesses beyond those observed in the LLM runs. The searches find split-order schedules that evade the implemented conduct rules; a learned baseline catches all 2,400 tested searched episodes but also flags 15\% of a large maker's sessions. Aggregate caps reduce the response-dependent gain unevenly, with smaller reductions when only near-touch depth affects prices. A response fitted to five Nasdaq stocks and a separate ABIDES experiment expose the dependence of these estimates on market assumptions. A distinct adversarial prompt elicited refusals from Sonnet in all 240 rounds, yielding no candidate strategies to evaluate. TradeGauntlet provides reproducible tests of controls around trading agents; the experiments do not establish legal compliance or real-market detection accuracy.
\end{abstract}

\keywords{agentic compliance \and autonomous trading agents \and market surveillance \and adversarial evaluation \and LLM agents}

\section{Introduction}\label{sec:intro}

An autonomous trading agent can receive a legitimate objective and still place orders that trigger a conduct rule. A broker therefore needs to test both the agent and the controls around it. We study agentic compliance at the point where an agent's decisions become market orders: which controls prevent problematic order behavior, and what do those controls cost the agent's intended task? The question is relevant as brokers allow agents to place and cancel orders on a customer's behalf~\cite{r23,r24}.

Two kinds of control are involved. A \emph{gate} checks an order before it reaches the market and may reject it. A \emph{detector}, or surveillance rule, inspects the order stream afterwards and raises an alert when it sees a specified pattern. Their effects can differ. A gate may reject a 60-lot order yet accept six 10-lot orders at the same price because it checks each order separately. A cap on total resting quantity closes that gap, but its value depends on how the market responds to displayed orders and on how much legitimate trading the cap prevents.

We built TradeGauntlet to measure \emph{detection} (conduct and baseline alerts), \emph{prevention} (the response-dependent gain remaining under a control), and \emph{cost} (execution quality and false alerts) separately. The simulator compares controls on specified market paths and records the attempted orders, rejections, fills, and alerts for replay. LLM agents pursue legitimate execution and market-making tasks. Scripted probes and searching adversaries then test weaknesses that those agents may never encounter. The LLM study covers 1,440 Haiku sessions and 240 Sonnet sessions; 1,342 and 220, respectively, were complete enough for the paired analysis.

The LLM results show why these distinctions matter. All conduct alerts were self-matches, which the gates block by construction. Sonnet's aggregate-gate condition also reduced two-sided fills by about 15\%, mainly through restrictions other than the aggregate cap. A conduct reminder reduced Sonnet's alerts but cost Haiku fills without a detectable alert benefit (Section~\ref{sec:llm}). The searched strategies expose a different weakness: they evade the implemented conduct rules, while a learned baseline catches the tested family at a cost of false alerts on a legitimate maker (Section~\ref{sec:baselines}). Neither result establishes how well the controls would perform against a broader population of agents.

The contribution is a reproducible workflow for testing compliance controls around autonomous trading agents, supported by the LLM comparison, two evaluated search methods (random search and a policy gradient), and scripted mechanism checks. We fit a price response to five Nasdaq stocks, test its shape in the separate ABIDES simulator, and examine sensitivity to response shape and real price paths. The economic findings remain conditional on those assumptions. A separate attempt to use an LLM as the adversary yielded only refusals and no evaluated candidates. All trading is simulated, and no result measures the prevalence of unlawful conduct in real markets.

\section{Related work}

\emph{Spoofing and its economics.} Agent-based studies show how spoofing works and how markets respond to it~\cite{r1,r7}. Others model its trade-off with fines~\cite{r18}, measure it in Korean market data~\cite{r34}, show when it arises in equilibrium~\cite{r35}, or cloak part of the book~\cite{r36}, an intervention that, like our caps, limits how much displayed depth moves prices. Tao et al.~\cite{r19} and Chen, Hsieh, and Yang~\cite{r20} bring TMX Level-2 and Taiwan derivatives data to detection and price banding, and Do and Putni\c{n}\v{s}~\cite{r13} and Fabre and Challet~\cite{r10} use prosecuted cases and Level-3 data. That is evidence our design lacks: we fit a response to Nasdaq data but have no external labels of manipulation.

\emph{Adaptive traders and detectors.} Byrd~\cite{r2} trains an RL trader that discovers spoofing and a recognizer that discourages it, Wang and Wellman~\cite{r3} model adaptation between manipulation and detection, and Cartea, Chang, and Garc\'{i}a-Arenas~\cite{r12} and Gu et al.~\cite{r17} study manipulation by learning algorithms and across liquidity conditions. Tuccella, Nadler, and \c{S}erban~\cite{r37} train a recurrent spoofing detector. ABIDES~\cite{r4} supports richer agent interaction, and we use it to check our price response. Our adversaries are simpler: a random search and a policy gradient over a fixed family of split-order schedules, which keeps each strategy they find inspectable and replayable. Realism metrics~\cite{r38} and generative simulators~\cite{r39} test whether a simulator reproduces market stylized facts; our minimal book does not, so we also replay real price paths.

\emph{LLM agents and controls.} LLM studies evaluate judgments about market abuse~\cite{r25}, adherence to risk mandates~\cite{r26}, and trading across tasks and tools~\cite{r27,r28,r29}. ReguSim~\cite{r8} separates a trader's reasoning, orders, enforcement, and monitoring, but uses exogenous prices and fully executes accepted orders; Borjigin and He~\cite{r9} propose a runtime shield; and Kurshan, Balch, and Byrd~\cite{r5} and Byrd~\cite{r11} discuss agent-based governance and LLM-generated social posts. We instead compare interacting controls at the order level, on matched market paths, with scripted, searched, and LLM traders.

\section{TradeGauntlet: simulator, controls, and measurements}

Some terms recur throughout. The \emph{subject} is the account under test; everything else in the book is background liquidity or noise. The \emph{touch} is the best bid or ask. \emph{Reference depth} is the median external quantity at the two touches, excluding the subject's orders. Prices are integer \emph{ticks}, and wealth is measured in \emph{tick-lots}, one tick of price on one lot. A \emph{session} or \emph{episode} is one simulated trading period on one market \emph{path}, the sequence of exogenous price and noise draws fixed by a seed.

\subsection{Market mechanics}\label{sec:market}

TradeGauntlet uses its own matching engine, written for this project rather than built on ABIDES, with integer prices, price/time priority, and partial fills. Scenario runners fix the configuration and random seeds, run paired control conditions, record attempted and accepted actions, and replay saved attempts to check event hashes and outcomes. The runners can vary cap multiples, response thresholds, throttle windows, session lengths, and quoting cadences. Orders execute only inside the simulator; there is no broker or exchange connection. Table~\ref{tab:spec} in Appendix~\ref{sec:supp} lists the configuration.

The fundamental and noise streams are generated before trading starts. For a given seed and session length, every gate and response condition therefore receives the same exogenous draws. Unfilled market quantity expires. Background quotes are replaced between successive subject actions, so splitting an order also changes how long the display remains in the book.

The price response is driven by order-book imbalance. Immediately before each replenishment, let $B$ and $A$ be the total remaining buy and sell quantity, including the subject's. The imbalance is $I = (B - A)/\max(1, B + A)$. Under the \emph{step} response, background quotes shift up by two ticks when $I$ exceeds a threshold $t$, down by two ticks when $I$ is below $-t$, and otherwise stay where they are. The default threshold is 0.4; Appendix~\ref{sec:appendix} varies it. In the \emph{unresponsive} condition the quotes never shift. The step response is stipulated, not estimated from market data.

Because every result under the step response hinges on a threshold we chose, we add a second response with no threshold. Under the \emph{linear} response, background quotes move $\kappa$ ticks in the direction of $I$ with probability $|I|$ and otherwise stay centred on the fundamental, so the expected displacement is $\kappa I$ ticks. This follows two empirical regularities: the probability that the next price move is upward rises with order-book imbalance, roughly linearly for large-tick stocks~\cite{r31}, and short-horizon price changes are approximately linear in order-flow imbalance~\cite{r30}. We use $\kappa = 1$ as a stress setting. The uniform draws that decide each move come from a separate seeded stream generated before trading, so all gate conditions on a seed share them. Section~\ref{sec:calibration} fits the move probability to market data.

\subsection{Pre-trade gates}

Every condition enforces a bound on the account's worst-case position. The \emph{per-order gate} (called legacy in the code) also rejects a limit order that is larger than twice reference depth, that would occupy a third resting price level on one side, or that would trade against one of the account's own resting orders (a self-match). The \emph{aggregate gate} adds a cap per side: the subject's remaining limit quantity plus the full proposed quantity must not exceed $m$ times reference depth. The default is $m = 2$. The sensitivity checks vary this aggregate multiple but keep the per-order limit at twice reference depth.

Both gates can also throttle the account: they reject new limit and market orders when $C/N > 0.9$ and $N \geq 20$, where $C$ counts the subject's cancellations and $N$ its accepted submissions in a trailing window of 1,800 seconds by default. Rejected attempts do not count towards $N$, and cancellations are always allowed. To switch the throttle off, we make the minimum count unreachable and leave the other checks unchanged. The aggregate check runs only at submission, so it does not revisit resting orders when external depth changes. It also checks a marketable limit order at its full quantity, which makes its rejection cost deliberately conservative.

\subsection{Pattern definitions}\label{sec:patterns}

For a canceled order $i$, let $q_i$ be submitted quantity, $d_i$ its submission-time external reference depth, and $a_i$ its age at cancellation. Let $h_i$ be the fraction of $q_i$ filled before cancellation and $z_i$ its distance from the same-side touch at submission. For a subject fill $f$ on the opposite side, define $Q(i,f)$ to hold when $i$ was nonmarketable at submission, submission precedes $f$, $f$ precedes cancellation, $a_i \leq W$, and $h_i < 0.1$. Event sequence numbers determine ordering within a timestamp. Table~\ref{tab:patterns} lists the implemented rules.

\begin{table}[htbp]
\centering
\small
\begin{tabularx}{\linewidth}{lX}
\toprule
\textbf{Signal} & \textbf{Condition for an alert} \\
\midrule
Individual pattern & Some $i$ and $f$ satisfy $Q(i,f)$, $q_i \geq 3d_i$, and $z_i \leq 3$ ticks. \\
Layer pattern & For one $f$ and one displayed side, qualifying $Q(i,f)$ orders occupy at least three distinct prices. No individual-size or distance test is added. \\
Aggregate pattern & For one $f$ and displayed side, at least two $Q(i,f)$ orders have $z_i \leq 3$ and positive remaining quantity $r_i(f)$. Their sum is at least $3\max_i d_i$, three times the largest reference depth in that group. \\
Self-match & The subject is both buyer and seller in a fill. \\
\bottomrule
\end{tabularx}
\caption{Conduct-alert patterns. Their union is the conduct-alert signal.}
\label{tab:patterns}
\end{table}

Remaining quantity $r_i(f)$ is measured immediately before the shared fill, after any earlier partial executions. The \emph{conduct alert} is the union of these four signals; a separate cancellation-ratio diagnostic is not part of it. The scoring window is $W = 180$ seconds by default, shared by the individual, layer, and aggregate patterns. Table~\ref{tab:split} keeps the 120-second window of the first split-order comparison, and Appendix~\ref{sec:cadence} varies $W$. These patterns are not a complete legal test and do not estimate intent.

\subsection{The three measurements}\label{sec:design}

We score every control on three measurements. Within each experiment, control contrasts use the same market paths; Section~\ref{sec:prevention} separately identifies a descriptive comparison of maxima across search experiments with different test seeds. \emph{Detection} is the share of episodes with a conduct alert (Table~\ref{tab:patterns}) or, in Section~\ref{sec:baselines}, a baseline alert. \emph{Prevention} is measured by the gain a strategy draws from the price response, which a gate may or may not remove. \emph{Cost} is measured through task completion and two-sided fills for LLM agents, and through fills, implementation shortfall, P\&L, edge, markout, and false alerts for legitimate scripts.

Terminal wealth $V$ is cash plus inventory valued at the terminal fundamental, after fees. For gate $g$, parameters $p$, and seed $s$, the response-dependent difference is
\[ D(g,p,s) = V_{\mathrm{resp}}(g,p,s) - V_{\mathrm{unresp}}(g,p,s). \]
Both conditions use the same gate and exogenous path. Positive $D$ means the strategy benefits from the price response. It is not a legal or welfare measure, and it is not the strategy's profit: a schedule can have $D > 0$ and still lose money overall. Executed lots, rejected actions by reason, and response activations are reported separately.

The scripted \emph{split probe} submits six 10-lot orders at one price in each cycle, executes two lots on the opposite side, and cancels whatever display survives; successive cycles alternate sides. The legitimate controls are a small and a large two-sided market maker, a patient maker, a passive scheduled executor, and a 10-lot slice executor. We assigned these scripts their objectives, and the assignments have not been independently reviewed. The scheduled executor is named \texttt{vwap\_executor} in the code but uses neither a volume forecast nor a VWAP benchmark.

\section{LLM agents under the same controls}\label{sec:llm}

We first evaluate two language-model agents given legitimate tasks: do their orders trigger conduct alerts, do gates prevent those events, and what do the controls cost in task performance? We ran Claude Haiku 5.5 and Claude Sonnet 5.5 as the subject trader (\texttt{agent\_study}). Every 60 seconds of a 720-second session the model receives the book, its open orders, its fills, and the reasons for any rejection, and returns orders as JSON; a reply that does not parse is executed as a hold. The model makes fresh decisions in each condition, allowing it to respond to the gate's feedback.

The design crosses four factors: the \emph{task} (buy a 40-lot parent order at low cost, or make a two-sided market with absolute inventory of at most 30 lots), the \emph{prompt} (the task alone, or plus a one-sentence conduct reminder), the \emph{gate} (none, per-order, or per-order plus the twice-depth cap), and the \emph{price response} (off, or a step or linear response). Market paths come from three held-out families: liquidity withdrawal, trend reversal, and a price jump. Haiku ran the full grid on 10 seeds per family (1,440 episodes), and Sonnet ran no gate and the aggregate gate with the step response on 5 seeds per family (240 episodes).

Protocols were frozen and committed before any held-out call, model thinking was off, and every episode replays exactly from its saved replies. Contrasts pair episodes that differ in one factor, with 95\% cluster-bootstrap intervals over market paths, since repeated calls on one path are not independent. The study cost 18.60 USD at list prices.

\begin{table}[t]
\centering
\caption{LLM market makers on held-out paths, pooled over families, seeds, and response settings. Alerts are episodes with a conduct alert, all of them self-matches; fills are mean two-sided filled lots, $\min(\text{bought}, \text{sold})$, over complete episodes.}\label{tab:llm}
\begin{tabular}{llrrr}
\toprule
Model & Prompt, gate & Episodes (complete) & Alerts & Fills \\
\midrule
Haiku 5.5 & task only, off & 120 (120) & 2 & 13.3 \\
 & task only, per-order & 120 (118) & 0 & 13.1 \\
 & task only, aggregate & 120 (118) & 0 & 14.0 \\
 & reminder, off & 120 (102) & 1 & 10.9 \\
 & reminder, per-order & 120 (106) & 0 & 10.4 \\
 & reminder, aggregate & 120 (107) & 0 & 10.5 \\
\midrule
Sonnet 5.5 & task only, off & 30 (29) & 10 & 22.9 \\
 & task only, aggregate & 30 (30) & 0 & 18.1 \\
 & reminder, off & 30 (28) & 2 & 20.4 \\
 & reminder, aggregate & 30 (30) & 0 & 18.7 \\
\bottomrule
\end{tabular}
\end{table}

On the buy task, neither model triggered a conduct alert in any arm. Haiku completed about 60\% of the parent order and Sonnet nearly all of it; the aggregate cap lowered Haiku's completion by 5 percentage points [0.5, 10].

In market making, every conduct alert in the study is a self-match: a new order that crossed the model's own resting quote (Table~\ref{tab:llm}). Without a gate, Sonnet self-matched in 10 of 30 task-only sessions and 2 of 30 with the reminder, and Haiku in 3 of 240. Both gates reduce this to zero, lowering Sonnet's alert rate by 0.19 [0.09, 0.31]. This illustrates the coupling formalized in Observation~\ref{prop:coupling}: the gates reject any order that would trade against the account's own resting order (16 such rejections for Sonnet), so the self-match detector cannot fire on gated traffic.

The informative contrasts are the cost and the prompt. The aggregate gate cost Sonnet 3.2 [1.9, 4.3] lots of two-sided fills, about 15\% of its ungated fills. Of its 89 rejections, 63 came from the per-order layer limit, as it quoted several price levels, 16 from self-match prevention, and only 9 from the aggregate cap. This matches Section~\ref{sec:cost}, where the cost also sits outside the aggregate cap. The reminder lowered Sonnet's alert rate by 0.11 [0.02, 0.20] with no detectable change in fills ($-0.9$ [$-2.2$, 0.4]). For Haiku, which rarely self-matched, the reminder cost 2.8 [2.0, 3.7] lots of fills and removed no detectable alerts. The price response lowered Haiku's buy completion by 3 points [0.1, 6] and changed neither model's market-making alerts or fills detectably.

Episodes with an unparseable reply (98 of 1,440 for Haiku, 20 of 240 for Sonnet) are reported as incomplete and their pairs are dropped.

\paragraph{An LLM as the adversary.} We also asked Sonnet 5.5 to take the search's place (\texttt{paper\_study.llm\_adversary}), with a system prompt describing the simulator, the response, the gates, and the exact conduct rules, the objective of Section~\ref{sec:adversary}, and the random search's budget of 80 evaluations per cell, shown each round with every result so far. Across 12 cells it declined in all 240 rounds, saying that the task asks for a spoofing schedule tuned to evade surveillance and offering help with detection instead (cost 1.69 USD). We did not reword the prompt. This concerns one model and one prompt, and a model that refuses does not limit what a human or a search can find.

\section{The price response}\label{sec:calibration}

Prevention is measured through the gain from moving the price, so the result depends on how displayed depth moves prices. We fit the response to market data and then check its shape in an agent-based market where no response is imposed.

\subsection{Fitting to Nasdaq data}

We use the five free LOBSTER samples: AAPL, AMZN, GOOG, INTC, and MSFT on 21 June 2012, reconstructed from Nasdaq messages at 10 levels (148,000 to 669,000 book updates per stock). AAPL, AMZN, and GOOG traded between about \$220 and \$580 with spreads of several ticks. INTC and MSFT traded near \$27 and \$30 and usually had a one-tick spread, so they are large-tick stocks in the sense of~\cite{r31}.

On a one-second grid, excluding the first and last ten minutes of the day, we regress the mid-price change over the next $h$ seconds, in units of the prevailing half-spread, on the imbalance of the top three levels. Three levels match the simulator's three background levels. Intervals use a block bootstrap with five-minute blocks.

The fitted slope must be mapped into simulator units, and there are two reasonable ways to do it. The simulator's half-spread is one tick. If its spread stands for the market's spread, the one-second slope is directly comparable with $\kappa s$, and we set the fitted scale $s$ to a stock's one-second point estimate; this is the \emph{half-spread mapping}, used unless stated otherwise. If instead a simulator tick stands for a market tick, the comparable scale is the slope times the median half-spread in ticks (the \emph{tick mapping}). Section~\ref{sec:adversary} reports results under both. AMZN and GOOG give slopes close to AAPL's, so the adversary uses AAPL, INTC, and MSFT as its fitted settings.

\begin{table}[htbp]
\centering
\footnotesize
\setlength{\tabcolsep}{2.5pt}
\begin{tabular}{lrrll}
\toprule
\textbf{Stock} & \textbf{Spread (ticks)} & \textbf{Obs.} & \textbf{Top three levels, 1 s} & \textbf{Best level, 1 s} \\
\midrule
AAPL & 15 & 18,028 & 0.018 [$-$0.003, 0.039] & 0.052 [0.031, 0.075] \\
AMZN & 12 & 15,340 & 0.032 [0.015, 0.049] & 0.036 [0.022, 0.050] \\
GOOG & 26 & 11,152 & 0.016 [$-$0.005, 0.037] & 0.025 [0.007, 0.044] \\
\midrule
INTC & 1 & 18,889 & 0.348 [0.287, 0.417] & 0.201 [0.172, 0.229] \\
MSFT & 1 & 19,220 & 0.479 [0.406, 0.574] & 0.252 [0.222, 0.284] \\
\bottomrule
\end{tabular}
\caption{Slope of the future mid-price change, in half-spreads, on order-book imbalance, 21 June 2012. Spread is the median quoted spread; observations are one-second grid points. Brackets are 95\% block-bootstrap intervals. Slopes at 5 and 10 seconds, quoted in the text, and at other horizons are saved with the results. Produced by \texttt{paper\_study.calibration}.}
\label{tab:calibration}
\end{table}

Under the half-spread mapping, the stocks split by tick size. For the three wide-spread stocks, the top-three-level slope is at most 0.032 half-spreads per unit of imbalance at one second and at most 0.069 at any horizon; for AAPL the sign even reverses at five and ten seconds. For INTC and MSFT, imbalance predicts the next move strongly: the one-second slope is 0.35 and 0.48 half-spreads per unit of imbalance, about a third to a half of the $\kappa = 1$ stress setting, and it grows to about two half-spreads at ten seconds. The simulator applies its response every second, so we use the one-second slopes; the longer horizons are not comparable with a per-second response.

Measured in market ticks instead, the one-second slopes are close for all five stocks, from 0.13 (AAPL) to 0.24 (MSFT) ticks per unit of imbalance, so the contrast between the two groups depends on the unit. Gould and Bonart~\cite{r31} study the best-level queue; for the one-tick stocks our top-three slope is about twice the best-level slope (Table~\ref{tab:calibration}). All of these are predictive regressions on five stocks on one day. Displayed orders that are not meant to trade may move prices less than organic depth does, and the regressions do not identify the causal effect of one trader's orders.

\subsection{Checking the shape in ABIDES}\label{sec:abides}

TradeGauntlet imposes its price response. ABIDES~\cite{r4} does not: in its RMSC04 reference market, 1,116 noise, value, momentum, and adaptive market-making agents trade through an exchange agent, and prices move only through their reactions. We precompute the ABIDES fundamental on a one-second grid before trading, so every arm of a seed shares the same fundamental path. A first check (\texttt{paper\_study.abides\_check}) added an agent that rests a display of $c$ times reference depth one tick behind the touch every 60 seconds, executes 20\% of reference depth two seconds later, and cancels. Over 100 seeds, its wealth gain from displaying is $-0.54$ to $-0.20$ cents per executed share for $c = 1$, 2, and 6, with intervals that include zero, but the minimum detectable effect, about 2 cents per share, is ten to twenty times the gains the fitted responses give in TradeGauntlet. Wealth is too noisy for this question.

\paragraph{Does the display move the price?} A second check measures the mechanism directly (\texttt{paper\_study.abides\_response}). Every 60 seconds a probe agent rests a display of $c$ times reference depth as six orders, alternating sides, reads the midpoint 1, 2, and 5 seconds later, and cancels without executing. The effect at each horizon is the midpoint's move in the display's direction minus the same move in a $c = 0$ arm on the same seed and cycle, in median half-spreads. Displays of $c = 1$, 2, or 6 rest one tick behind the touch; a fourth, $c = 2$ five ticks behind, is rarely reached by trading and so separates agents reading the book from displayed orders absorbing trades. Each arm runs on 100 seeds in RMSC04 and in a thinner book with 20 instead of 102 value agents (2,600 to 2,800 paired cycles per arm), and Table~\ref{tab:abidesresponse} compares the effect with what TradeGauntlet's fits predict for the imbalance the display creates.

\begin{table}[t]
\centering
\caption{Effect of the probe's display on the ABIDES midpoint, in half-spreads in the display's direction, against the 1-second effect TradeGauntlet's MSFT and INTC fits predict for the same change in imbalance $\Delta I$. Brackets: 95\% intervals from 4,000 bootstrap resamples of the 100 seeds.}
\label{tab:abidesresponse}
\small
\begin{tabular}{llrrrrr}
\toprule
Market & Display & $\Delta I$ & 1\,s & 5\,s & MSFT fit & INTC fit \\
\midrule
RMSC04 & $c=1$, 1 tick & 0.15 & 0.09 [0.05, 0.13] & 0.46 [0.33, 0.60] & 0.07 & 0.05 \\
 & $c=2$, 1 tick & 0.26 & 0.11 [0.06, 0.15] & 0.49 [0.36, 0.63] & 0.12 & 0.09 \\
 & $c=6$, 1 tick & 0.50 & 0.09 [0.05, 0.13] & 0.53 [0.41, 0.67] & 0.24 & 0.17 \\
 & $c=2$, 5 ticks & 0.26 & 0.01 [$-0.02$, 0.04] & 0.16 [0.07, 0.26] & 0.13 & 0.09 \\
Thin book & $c=1$, 1 tick & 0.15 & 0.07 [0.02, 0.13] & 0.34 [0.18, 0.52] & 0.07 & 0.05 \\
 & $c=2$, 1 tick & 0.26 & 0.06 [0.01, 0.12] & 0.36 [0.20, 0.54] & 0.12 & 0.09 \\
 & $c=6$, 1 tick & 0.49 & 0.06 [0.00, 0.11] & 0.40 [0.23, 0.60] & 0.24 & 0.17 \\
 & $c=2$, 5 ticks & 0.26 & 0.04 [0.00, 0.09] & 0.23 [0.10, 0.39] & 0.12 & 0.09 \\
\bottomrule
\end{tabular}
\end{table}

Three findings follow. \emph{First, an agent's own display does move the price in ABIDES.} One tick behind the touch, every display moves the midpoint in its direction at every horizon, and at $c = 1$ and $c = 2$ the 1-second effect is about what the MSFT and INTC fits predict. The fitted slopes are therefore not an implausible size for small displays near the touch. \emph{Second, the effect does not grow with size.} From $c = 1$ to $c = 6$ the induced imbalance triples but the 1-second effect stays near 0.09 half-spreads in RMSC04, so the effect per unit of imbalance falls from 0.59 to 0.18, and at $c = 6$ the intervals exclude both fitted predictions. The linear and fitted responses, which scale with imbalance, overstate what large displays earn. \emph{Third, depth away from the touch barely moves the price.} Five ticks back, the same display moves the RMSC04 midpoint by 0.01 half-spreads at 1 second, with an interval that excludes both fitted predictions, and in the thin book the interval excludes the MSFT prediction. Displays near the touch are filled 5 to 9 lots per cycle in RMSC04 against fewer than 1 five ticks back, so much of the near-touch effect may come from displayed orders absorbing trades rather than from agents reading the book. By 5 seconds the far display has an effect too, smaller than the near ones.

\section{Zero alerts can be built in}\label{sec:aggregation}

The first thing an alert count can hide is that the count was fixed before any trading took place.

\begin{table}[htbp]
\centering
\small
\begin{tabular}{lllll}
\toprule
\textbf{Gate} & \textbf{Original alerts} & \textbf{With aggregate signal} & \textbf{Executed lots} & \textbf{Wealth} \\
\midrule
off & 0/40 & 40/40 & 24.00 [24.00, 24.00] & 22.05 [14.80, 32.80] \\
legacy & 0/40 & 40/40 & 24.00 [24.00, 24.00] & 22.05 [14.80, 32.80] \\
aggregate & 0/40 & 0/40 & 24.00 [24.00, 24.00] & -25.95 [-33.20, -15.20] \\
\bottomrule
\end{tabular}
\caption{Six-by-ten split probe, \texttt{pilot.review\_study}, seeds 3000-3039, responsive market. Both gates include the default throttle. The 120-second scoring window is retained from this comparison; rescoring at the current 180-second default leaves these alert counts unchanged. Alerts count episodes; volume and wealth are mean [minimum, maximum]. Wealth is in tick-lots.}
\label{tab:split}
\end{table}

Table~\ref{tab:split} runs the split probe on 40 paths. Each split order passes the individual-size check, and all of them sit at one price level, so the individual and layer detectors flag none of the 40 paths. The aggregate detector flags the sequence whenever it is admitted. Under the aggregate gate, however, the absence of an aggregate alert says nothing about the detector. It follows from the inequality the gate and detector share.

\begin{proposition}[Gate--detector coupling]\label{prop:coupling}
Suppose every order in a candidate group was admitted by the aggregate gate with multiple $m$, and the aggregate pattern alerts when the group's remaining quantity at the shared fill is at least $k \max_i d_i$. If $m < k$, the aggregate pattern never alerts on that group.
\end{proposition}

\begin{proof}
Let $j$ be the last group order admitted. At its admission the gate requires that the remaining same-side quantity, including $q_j$, be at most $m d_j \le m \max_i d_i$. Orders admitted before $j$ can only lose quantity afterwards, through fills or cancellation, and no order grows after admission, so the remaining quantity at the shared fill is also at most $m \max_i d_i < k \max_i d_i$.
\end{proof}

With $k = 3$ and the default $m = 2$, the zero alert count under the aggregate gate is guaranteed and says nothing about how accurate the detector is. When the gate keeps the tested quantity strictly below the detector's alert threshold, the absence of that alert measures the gate's restriction.

The same coupling appears with LLM traders (Section~\ref{sec:llm}): the only conduct alerts they raised were self-matches, and every gate rejects an order that would trade against the account's own resting order.

\section{Detection: a searching adversary evades the rules}\label{sec:adversary}

The split probe is a script we wrote, and the step response uses a threshold we chose. This section replaces both: a search chooses the strategy, and the response is linear or fitted to data. We ask whether the conduct rules detect what the search finds; Section~\ref{sec:prevention} asks how much of its gain the gates remove. The runner is \texttt{paper\_study.adversary}. Its plan, search logs, held-out episodes, and summary are saved, and a verify command reruns every held-out evaluation and reproduces the summary exactly.

\paragraph{Adversary.} The adversary chooses a split-order schedule from a family with six parameters: the number of orders (2 to 8), total displayed quantity (10 to 80 lots), how far behind the touch the orders rest (0 to 5 ticks), the gaps between submissions, how the orders' lifetimes overlap (full, staggered, or disjoint), and the delay before a 2-lot execution on the opposite side. Each candidate is scored on 10 training paths by its mean $D$ minus 100 times its fraction of episodes with a conduct alert, so any schedule prone to alerts loses. The search draws 40 random candidates, then makes 40 single-parameter local moves, keeping each improvement. The selected schedule is then scored on 100 held-out paths that the search never saw. We run a separate search for each of four controls (no gate, the per-order gate, and the aggregate cap at $m = 2$ and at $m = 1$) and each of six responses, with the cancellation throttle on: 24 searches in all.

\paragraph{Responses.} The six responses are the step response at $t = 0.4$ and $t = 0.2$, the linear response with $\kappa = 1$ as a stress setting, and three fitted settings in which the move probability is $s|I|$, with $s$ estimated for one stock (Section~\ref{sec:calibration}).

\begin{table}[htbp]
\centering
\footnotesize
\setlength{\tabcolsep}{4pt}
\begin{tabular}{lllll}
\toprule
\textbf{Response} & \textbf{No gate} & \textbf{Per-order} & \textbf{Aggregate, $m = 2$} & \textbf{Aggregate, $m = 1$} \\
\midrule
Step, $t = 0.4$ & 0.00 [0.00, 0.00] & 0.00 [0.00, 0.00] & 0.00 [0.00, 0.00] & 0.00 [0.00, 0.00] \\
Step, $t = 0.2$ & 48.00 [48.00, 48.00] & 48.00 [48.00, 48.00] & 48.00 [48.00, 48.00] & 0.00 [0.00, 0.00] \\
Linear, $\kappa = 1$ & 9.78 [9.26, 10.30] & 8.24 [7.62, 8.88] & 6.14 [5.60, 6.70] & 3.32 [2.86, 3.80] \\
Fitted, MSFT & 2.84 [2.42, 3.28] & 3.14 [2.68, 3.60] & 2.68 [2.30, 3.06] & 1.37 [0.97, 1.76] \\
Fitted, INTC & 2.00 [1.64, 2.38] & 2.68 [2.26, 3.12] & 2.00 [1.62, 2.38] & 1.10 [0.82, 1.38] \\
Fitted, AAPL & 0.00 [0.00, 0.00] & 0.12 [0.04, 0.22] & 0.04 [0.00, 0.10] & 0.04 [0.00, 0.10] \\
\bottomrule
\end{tabular}
\caption{Held-out mean $D$, in tick-lots per session, of the schedule the search selected for each control and response, over 100 held-out paths (seeds 9100-9199). Brackets are 95\% bootstrap intervals. No held-out episode in any cell raised a conduct alert. Training paths are seeds 9000-9009; the throttle is on.}
\label{tab:adversary}
\end{table}

\emph{The detectors do not stop a searching adversary.} With no gate and the $\kappa = 1$ response, the fixed six-by-ten probe earns 12.06 tick-lots per session (95\% CI 11.38 to 12.72) but alerts on all 100 held-out paths. The searched schedule keeps 9.78, about 81\% of that gain, with no alerts. It places two 40-lot orders, one and five ticks behind the touch, and cancels the first before submitting the second, so the only order live at the execution sits outside the detectors' three-tick distance condition. The same holds under the responses fitted to the one-tick stocks: the fixed probe earns 4.32 (INTC) and 5.86 (MSFT) with an alert on every path, and the search finds alert-free schedules earning 2.00 and 2.84.

\paragraph{Robustness.} We fixed four further checks in one plan before running them (\texttt{paper\_study.robustness}): restarting the search, limiting the response to depth near the touch, changing the unit mapping, and separating the display gain from $D$. Each cell repeats its search with three independent search seeds at the original budget and keeps the restart with the best training score. Table~\ref{tab:robustness} in Appendix~\ref{sec:supp} reports that schedule's held-out gain on seeds 9100-9199 and the range across the three restarts. The restarts bear on prevention and are discussed in Section~\ref{sec:prevention}, and the display gain in Section~\ref{sec:cost}.

\emph{Response window.} In the default response, depth at any distance from the touch moves the price as much as depth at the touch, and the schedules above rest orders up to five ticks back, outside the fitted top three levels. When only orders within three ticks of each side's best price count, the search still finds alert-free schedules under every control: 7.30 tick-lots under $\kappa = 1$, 2.84 under the MSFT fit, and 2.42 under the INTC fit with no gate, with no conduct alert on any held-out path. In eight of these nine cells, the selected schedule rests all its orders one to three ticks behind the touch. The evasion therefore does not depend on depth outside the fitted window; the cap's effect does (Section~\ref{sec:prevention}).

\emph{The size of the gain depends on the unit mapping.} Under the half-spread mapping it follows the stock's tick size. Under the AAPL fit, which stands for the three wide-spread stocks, no searched schedule earns more than 0.12 tick-lots per session, and the fixed probe earns 0.50 (0.32 to 0.70) with no gate. Under the one-tick fits, the searched gains in Table~\ref{tab:adversary} are 20 to 30\% of the $\kappa = 1$ stress setting and positive on 66 to 80 of 100 held-out paths with no gate, the per-order gate, or the $m = 2$ cap. Under $m = 1$ they remain positive in this search, but not in every later one. The checks below show that this tick-size contrast is an artefact of the mapping.

\emph{Unit mapping.} The fitted scales map one simulator half-spread to one market half-spread. Under the tick mapping, the scale is the slope times the median half-spread in ticks: 0.13 for AAPL and 0.24 for MSFT, the two ends of the five stocks' range. The searched gains are then similar for the two stocks: 0.88 and 1.52 with no gate, and 0.96 and 1.12 under the $m = 2$ cap. The contrast between wide-spread and one-tick stocks in Table~\ref{tab:adversary} is therefore a consequence of the half-spread mapping. Under either mapping, imbalance moves prices enough for a searched schedule to gain about one to three tick-lots per session without an alert.

Selection on training paths overstates performance, and a modest search does not always find the best schedule. Under the INTC fit, for example, the per-order-gate cell exceeds the no-gate cell (2.68 against 2.00): the gate's rejections turn the selected schedule into one the no-gate search did not find. The searched gains are therefore lower bounds, and scoring on held-out paths is necessary rather than cosmetic.

\subsection{Real price paths}\label{sec:realpaths}

The simulator's fundamental moves by at most one tick every 30 seconds, so its prices lack the heavy tails and volatility clustering of real markets. Table~\ref{tab:facts} in Appendix~\ref{sec:supp} applies the same estimators to 720-second sessions of the simulator and to 30 consecutive 740-second windows of each LOBSTER sample, all in half-spread units (\texttt{paper\_study.empirical\_paths}). Real 60-second returns are two to four times as volatile as the simulator's, heavy-tailed, and show volatility clustering; in both, returns are serially uncorrelated.

We therefore replaced the fundamental with the real mid-price path of each window, rounded to whole half-spreads, kept the noise and response streams of seed $9100 + k$ for window $k$, and replayed the schedules selected on synthetic training paths, unchanged, with the response fitted to the same stock (synthetic-policy columns of Table~\ref{tab:realsearch}). The gains survive. Under the MSFT fit, $D$ is 2.67 with no gate and 2.60 under the twice-depth cap, against 2.84 and 2.68 on synthetic paths, and under $\kappa = 1$ it is 9.73 against 9.78. The evasion of the ungated schedules does not survive. On real paths they trigger alerts in 16 to 23 of 30 windows: self-matches under the fitted responses, and the spoofing pattern under $\kappa = 1$. Larger price moves bring the displayed orders into the path of the schedule's own executions, or within the spoofing distance. Under the caps, whose per-order checks include self-match prevention, no window alerts. With 30 windows the intervals are wide; under the cap at reference depth they include zero for INTC and run from 1.0 to 5.6 for MSFT.

These schedules were tuned on calm paths, so we also reran the search on the real paths (\texttt{paper\_study.real\_path\_search}, Table~\ref{tab:realsearch}). The 30 windows of each stock form three folds by window number modulo 3. For each fold, the search, with the same space, budget, and alert penalty as before, is trained on the other 20 windows, and its schedule is scored on the 10 windows in the fold. Every window therefore gets one out-of-sample score. The schedule selected on synthetic paths is scored on the same windows for a paired comparison.

With no gate, searching on real paths brings the alerts down from 17, 23, and 16 of 30 windows to 3, 3, and 4. The gain does not all survive. Under the MSFT fit it stays at 2.64 (0.70 to 6.05). Under $\kappa = 1$ it falls from 9.73 to 2.70, a paired difference of $-7.03$ ($-8.47$ to $-5.63$), so avoiding alerts on volatile paths costs about 70\% of the gain. Under the INTC fit the mean is $-5.98$ ($-21.95$ to 2.40), so no gain is detectable.

Under the gates, only three windows alert, all under the per-order gate with the MSFT fit. The per-order gate leaves 9.63 (1.60 to 22.23) under the MSFT fit and 3.80 (1.00 to 7.38) under the INTC fit. The twice-depth cap leaves 4.41 (0.33 to 11.18) and 1.93 (1.27 to 2.60), but under $\kappa = 1$ its interval includes zero. At reference depth the intervals include zero for both fitted responses. The schedules selected in different folds differ, and their training scores differ by more than tenfold in some cells, so these numbers, like those in Table~\ref{tab:adversary}, are lower bounds from an unstable search rather than estimates of what an optimal adversary would earn.

\begin{table}[t]
\centering
\caption{Search rerun on real mid-price windows (\texttt{results/real-path-search-v1}). Each of 30 windows is scored by a schedule searched on the other two folds (real-path policy), and by the schedule selected on synthetic paths and replayed unchanged (synthetic policy; its held-out synthetic-path gains are in Table~\ref{tab:adversary}). Mean $D$ in tick-lots per session with 95\% bootstrap interval over windows, and windows with a conduct alert.}\label{tab:realsearch}
\begin{tabular}{llrrrr}
\toprule
 & & \multicolumn{2}{c}{Real-path policy} & \multicolumn{2}{c}{Synthetic policy} \\
Response, path & Control & $D$ & Alerts & $D$ & Alerts \\
\midrule
MSFT fit, MSFT & none & 2.64 [0.70, 6.05] & 3/30 & 2.67 & 17/30 \\
 & per-order & 9.63 [1.60, 22.23] & 3/30 & 8.09 & 0/30 \\
 & $m = 2$ & 4.41 [0.33, 11.18] & 0/30 & 2.60 & 0/30 \\
 & $m = 1$ & $-0.17$ [$-2.54$, 1.24] & 0/30 & 3.05 & 0/30 \\
INTC fit, INTC & none & $-5.98$ [$-21.95$, 2.40] & 3/30 & 2.46 & 23/30 \\
 & per-order & 3.80 [1.00, 7.38] & 0/30 & 2.33 & 0/30 \\
 & $m = 2$ & 1.93 [1.27, 2.60] & 0/30 & 1.40 & 0/30 \\
 & $m = 1$ & 0.67 [$-0.28$, 1.46] & 0/30 & $-0.34$ & 0/30 \\
$\kappa = 1$, MSFT & none & 2.70 [1.84, 3.64] & 4/30 & 9.73 & 16/30 \\
 & per-order & 6.70 [$-1.29$, 18.94] & 0/30 & 7.22 & 4/30 \\
 & $m = 2$ & 1.60 [$-0.27$, 3.07] & 0/30 & 5.04 & 0/30 \\
 & $m = 1$ & 2.84 [1.97, 3.73] & 0/30 & 3.07 & 0/30 \\
\bottomrule
\end{tabular}
\end{table}

\subsection{Baseline detectors}\label{sec:baselines}

We wrote the conduct rules for this project, so their failure could simply reflect a weak rule. We therefore add two alternative detectors (\texttt{paper\_study.baselines}). The first is an experimental order-to-trade proxy, motivated by the venue controls in RTS 9~\cite{r33}: the number of submissions plus cancellations, divided by the greater of one and the account's fill count. We choose its threshold; this is not an implementation of the regulatory calculation. The second is a logistic-regression detector on seven per-session features of the order stream, following the detection-evasion setup of~\cite{r3}: the cancelled share of displayed quantity, the log order-to-trade ratio, peak display relative to reference depth, the share cancelled within 30 seconds of an own opposite-side fill, the share cancelled within 60 seconds of submission, the passive fill share, and the aggressor share. It is trained on 600 sessions of random schedules from the family and 50 sessions of the five legitimate scripts (seeds 9000-9009, MSFT-fitted response). Each threshold allows at most 5\% false alerts on the legitimate training sessions.

\begin{table}[htbp]
\centering
\small
\begin{tabular}{lrrrr}
\toprule
\textbf{Detector} & \textbf{Large maker} & \textbf{Other legitimate} & \textbf{Random schedules} & \textbf{Searched schedules} \\
\midrule
Conduct rules & 0/100 & 0/400 & 64/100 & 0/2,400 \\
Order-to-trade ratio & 22/100 & 1/400 & 99/100 & 1,816/2,400 \\
Learned & 15/100 & 0/400 & 100/100 & 2,400/2,400 \\
Any of the three & 23/100 & 1/400 & 100/100 & 2,400/2,400 \\
\bottomrule
\end{tabular}
\caption{Sessions that alert, on held-out seeds 9100-9199. The legitimate columns are false alerts; searched schedules are the 24 found against the conduct rules (Table~\ref{tab:adversary}), each on 100 paths. Produced by \texttt{paper\_study.baselines}.}
\label{tab:baselines}
\end{table}

The learned detector alone flags all 2,400 held-out sessions of the 24 schedules the adversary found. The ratio flags 1,816 (76\%), and none in five cells (the $m = 2$ cap under $\kappa = 1$ and the AAPL fit, and the $m = 1$ cap under both step responses and the INTC fit). Combining the detectors adds no detections of searched schedules on this test set, but increases false alerts on the large maker from 15 to 23 of 100 sessions.

A search that must evade all three detectors finds no alert-free schedule, but with larger executions per cycle its schedules earn several times the gains above (Appendix~\ref{sec:supp}). The gains reported for the conduct rules are therefore gains per two-lot execution, about 0.1 to 0.4 tick-lots per executed lot, not a ceiling on what displayed depth can earn.

The learned detector puts its largest weight on the aggressor share, because every searched schedule executes with a market order while the legitimate scripts trade passively. An adversary that executes passively, or mixes genuine quoting with its displays, could weaken that signal. The family contains no such strategies. The learned detector is also trained on members of the same schedule family, so its observed perfect recall is an in-distribution result, not evidence of general manipulation detection. Its 15\% false-alert rate on the large maker is a cost even within this restricted comparison.

\section{Prevention: what a cap removes depends on the adversary}\label{sec:prevention}

We use the reduction in response-dependent gain as the economic prevention endpoint, with the limits set out in Section~\ref{sec:design}. We compare the gain remaining under each control (Table~\ref{tab:adversary}) against searches of increasing strength. Every selected schedule in this section is conduct-alert-free on every held-out path, so that alert count alone does not distinguish the controls' economic effects.

\paragraph{Against the first search.} How much the aggregate cap protects depends on the response, and under the fitted responses it protects only partly. Under $\kappa = 1$, the cap at $m = 2$ cuts the searched gain from 9.78 to 6.14, and $m = 1$ cuts it to 3.32. Under the INTC and MSFT fits, $m = 2$ leaves the searched gain essentially unchanged in this search (2.00 and 2.68, against 2.00 and 2.84 with no gate), and $m = 1$ roughly halves it (1.10 and 1.37); the larger search below revises the MSFT comparison. The cap does remove the fixed probe's alerts, but with the throttle off the probe still earns 2.30 (INTC) and 3.04 (MSFT) under the cap, about what the search finds. Only the step responses give all-or-nothing protection (Appendix~\ref{sec:stepcheck}), and Observation~\ref{prop:boundary} predicts which step cells are zero.

\paragraph{Stronger searches.} A modest search can miss better schedules, so we ran two stronger ones on the nine cells behind these conclusions ($\kappa = 1$, INTC, and MSFT, each with no gate, $m = 2$, and $m = 1$). A random search with three times the budget, 20 training seeds, and 200 new held-out seeds (\texttt{paper\_study.adversary\_budget}) barely moves the $\kappa = 1$ results (9.98, 6.02, and 3.33), but under the fits it finds better ungated schedules: 5.03 (4.65 to 5.41) for MSFT, against 2.84, and 2.10 for INTC. A policy gradient (\texttt{paper\_study.rl\_adversary}), with one softmax distribution per schedule parameter updated by REINFORCE on batches of eight with standardized advantages, brings no advantage at the random search's budget of 80 samples: three runs per cell fall on both sides of it (Table~\ref{tab:rl} in Appendix~\ref{sec:supp}). With 800 samples it finds 4.86 (4.28 to 5.44) for MSFT and 3.32 (2.88 to 3.74) for INTC with no gate, but little more under the caps. Restarts of the original search differ by up to a factor of 1.6 (Table~\ref{tab:robustness}), as much as some differences between controls; the cap's reductions under $\kappa = 1$ exceed that range, but its effect at $m = 2$ under the fits in the first search does not. No selected schedule in any of these searches raised a conduct alert on any held-out path.

For a descriptive comparison across searches, the highest observed held-out gains at $m = 2$ are 2.84 against an ungated 5.03 for MSFT, 2.04 against 3.32 for INTC, and 6.24 against 9.98 under $\kappa = 1$. These ratios imply reductions of approximately 44\%, 39\%, and 37\%, respectively; at $m = 1$ the corresponding reductions are approximately 59 to 67\%. They apply to the default response using all displayed depth. We selected these maxima after inspecting the held-out scores, and the larger random search uses 200 new test seeds while the other searches use 100. The ratios are therefore comparisons of best observed scores, not paired estimates on a common untouched test set. Selection bias need not cancel between the gated and ungated arms. Estimating those reductions would require selecting schedules without using the final test set, then evaluating them on common fresh paths.

\paragraph{What the response shape implies.} The cap works by limiting display size, and the linear and fitted responses reward size in proportion to the imbalance it creates. In ABIDES, a display six times reference depth moves the price no more than one at reference depth, and depth five ticks back barely moves it (Section~\ref{sec:abides}). If real markets behave like ABIDES, the gains of large-display schedules, and with them the share a cap removes, are overstated. The response limited to depth near the touch, which ABIDES supports over the default, already gives the cap a smaller effect (Section~\ref{sec:adversary}): at $m = 2$ it removes 16\% of the $\kappa = 1$ gain, 24\% of the INTC gain, and none of the MSFT gain, while $m = 1$ removes roughly half. On real price paths, the per-order gate and the twice-depth cap leave a positive gain for both fitted stocks (Table~\ref{tab:realsearch}). ABIDES is itself a model, with one-tick spreads and agents whose rules we did not choose, so it bounds the shape of the response rather than calibrating it.

\section{Cost: legitimate traders gain from the response and pay for the controls}\label{sec:cost}

Controls act on legitimate traders in two ways. A legitimate trader may benefit from the same price response the adversary exploits, so a control that removes the gain also removes part of the legitimate trader's gain. And a control may block its trading or flag it.

\paragraph{What $D$ measures.} $D$ equals the display gain, a schedule's wealth minus that of the same schedule with its displayed orders removed, in every cell without the $m = 1$ cap. The schedules themselves lose money: total wealth is between $-15.7$ and $-24.5$ tick-lots per session in those cells, because they pay the spread on their market orders, and the price response reduces that loss rather than turning it into a profit.

Legitimate scripts gain from the response too. Under the MSFT fit, the patient maker's $D$ is 10.7 (4.2 to 17.5) with all depth and 20.0 (13.2 to 27.3) within three ticks, larger than any searched schedule's, and the large maker's is 8.8 (0.1 to 17.3) within three ticks; the executors and the small maker show no detectable gain. $D$ therefore measures how much a strategy benefits from the price response, not whether it manipulates. Our evaluation identifies the manipulative pattern through the conduct and baseline alerts and uses $D$ only to measure what a gate removes. A cap that removes the gain from displayed depth would also remove part of the patient maker's.

\paragraph{Execution quality.} Lost volume is the usual measure of a control's cost, but it can both understate and overstate the harm (Appendix~\ref{sec:throttle}). The passive executors lose no volume to the gate yet complete only about a third of their parent orders, and most of the large maker's volume loss comes from the cancellation throttle rather than the cap. Table~\ref{tab:execution} in Appendix~\ref{sec:supp} therefore scores the same scripts with standard execution-quality measures on 100 new seeds under the MSFT-fitted response.

The cap at $m = 2$ leaves both executors and the small maker unaffected: fill rates move by less than 0.1 percentage points and implementation shortfall by less than 0.003 ticks per lot. For the large maker, the throttle is what removes volume: with it, the maker loses 5.7 lots (9\%) and 2.5 tick-lots of P\&L ($-5.3$ to 0.3). With the cap alone its volume rises slightly (1.7 lots, 0.5 to 3.0) and its P\&L change is indistinguishable from zero. In both arms its edge per lot falls by 0.11 to 0.12 ticks while its 60-second markout improves slightly (by 0.02 to 0.03 ticks): the cap moves its quotes away from the best price, and the fills it keeps are slightly less adversely selected. The patient maker loses 31\% of its volume, but its P\&L change is indistinguishable from zero (1.4, $-7.3$ to 10.4) and its mean absolute inventory falls by 8.9 lots. A volume-based cost would rank this as the largest harm.

The stricter cap, $m = 1$, is costly wherever it binds. The scheduled executor fills 20 percentage points less of its parent order and the slice executor 8 points less, and both large makers are shut out almost entirely. It does not affect the small maker, and the executors' shortfall does not change detectably. That last result partly reflects the measure, which prices the unfilled remainder at the closing fundamental with no spread or impact and so understates the cost of under-execution. Under the step response the pattern is the same but the sizes differ: the throttled large maker loses 8.3 lots (14\%), and the patient maker's P\&L rises under the cap (7.2, 1.1 to 13.7).

\paragraph{False alerts.} Detection has a cost too. The conduct rules flag none of the legitimate sessions, but they also miss every searched schedule. The learned baseline catches all tested searched episodes and flags 15 of 100 held-out sessions of the large maker, which replaces its quotes often (Table~\ref{tab:baselines}). Combining detectors raises that false-alert count to 23 without improving recall on the searched family. These false alerts fall on the active legitimate maker that also bears a substantial throttle cost.

\section{Implications for agentic compliance and supervision}

For a team deploying trading agents, the workflow offers a way to compare a prompt, model, or control change on fixed scenarios before deployment. The useful record includes the attempted action, the gate's reason for accepting or rejecting it, the resulting fills, and the detector output. In our LLM study, that record distinguishes self-match prevention from an improvement in the agent's own choices, and identifies which checks account for lost fills. The simulator can support this testing process; it does not establish how a deployed agent will behave in a live market.

The gates correspond most directly to the pre-trade controls of SEC Rule 15c3-5~\cite{r32}, which does not prescribe our aggregate caps or throttles. FINRA Rule 3110~\cite{r15} establishes supervisory obligations. Notice 15-09~\cite{r14} identifies testing independent of code development as a suggested effective practice; our project-internal evaluation does not provide that independence. Rule 5210~\cite{r16} describes a disruptive-quoting pattern over several price levels that our layering pattern resembles only in part, and Notice 24-09~\cite{r6} explains that existing obligations apply to generative AI. The CFTC makes the intent to cancel before execution material~\cite{r21}; our order patterns do not establish that intent. EU RTS 6~\cite{r22} addresses testing methodology, surveillance adequacy, and replay of order data, motivating tests of both suspicious and legitimate activity. Our size multiples, ratios, and windows are experimental choices, not regulatory thresholds. An alert needs interpretation, and an accepted sequence is not a certification of compliance.

\section{Limitations}\label{sec:limitations}

\emph{Market.} The simulator has one account, one instrument, three replenishing levels with constant depth, and no queues calibrated to data, latency, strategic background traders, auctions, or cross-venue routing. Its own price path lacks the heavy tails and volatility clustering of the data~\cite{r38,r39}; real price paths, all from one day, address this only for the fundamental. The fitted response is a predictive regression on five stocks on one day in 2012, its strength depends on the unit mapping, and it does not identify the causal effect of a trader's own orders. ABIDES suggests that effect is close to the fit for small displays near the touch and smaller otherwise, but ABIDES is itself a model.

\emph{Strategies and detectors.} The strategies, labels, and detectors came from the same development process, and the adversaries search one six-parameter family, so their gains are lower bounds and the searches are unstable across restarts. Every schedule in the family executes with a market order, which the learned baseline exploits; passive and mixed strategies are untested. The detector window and the appendix settings were chosen after inspecting earlier results, so those comparisons are exploratory.

\emph{Costs and LLMs.} Sessions are short, the execution-quality benchmarks are the simulator's own prices, and the gate's conservative check of marketable limit orders shapes its measured cost. The LLM study covers two models from one provider, one wording per prompt, and market families authored in this project; 7 to 8\% of episodes are incomplete, which may bias the paired contrasts. The one adversarial LLM run ended in refusal, so the study says little about LLM adversaries in general.

\section{Conclusion and availability}

Agentic compliance requires testing the interaction between an agent's decisions and the controls that accept, reject, or flag them. In this study, gates prevented the LLM traders' self-matches but also changed execution, while a conduct reminder had different effects across the two models. Scripted and searched strategies exposed weaknesses outside the behavior observed in those LLM runs. The learned detector caught the tested search family, but its false alerts on a legitimate maker and reliance on the aggressor-share feature limit that result. Economic prevention depended on the search and on the assumed price response.

TradeGauntlet makes these interactions available as replayable tests. It can help an agent developer examine what a proposed control prevents, what passes through it, and what task performance it costs. Stronger evidence would require independently authored scenarios, more models and prompt wordings, passive and mixed adversaries, and a final common test set for policies selected across search methods. Fitting the response on more days and testing controls in richer markets would address the remaining gap between this simulator and deployment.

\subsection{Code and data availability}

Code, saved results, and reproduction instructions: \href{https://github.com/tokencraftdata/agentic-compliance}{github.com/tokencraftdata/agentic-compliance}. The experiment code and evidence used here are archived at commit \href{https://github.com/tokencraftdata/agentic-compliance/tree/a0d736fc6b6452e0d5ccd1d12a54e29e4c087ea9}{\texttt{a0d736f}}; this revision changes the presentation of those results.

\subsection{Use of generative AI tools}

Anthropic Claude Code and OpenAI Codex assisted with writing the simulation code, tests, and analysis scripts, with literature checks, and with drafting and editing the manuscript. Apart from the agent study in Section~\ref{sec:llm}, where Claude Haiku 5.5 and Sonnet 5.5 are the traders under study, no LLM was called in any reported experiment. The authors reviewed all generated material and are responsible for the methods, results, and references.

\appendix

\section{Supplementary results}\label{sec:supp}

\begin{figure}[t]
\centering
\includegraphics[width=\linewidth]{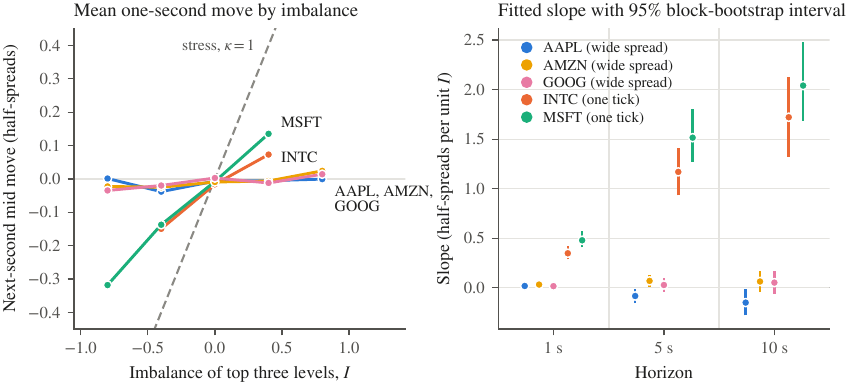}
\caption{Fitted price response. Left: mean next-second mid move by top-three-level imbalance, in five bins, omitting bins with fewer than 30 observations; the three wide-spread stocks overlap near zero. The dashed line is the $\kappa = 1$ stress setting. Right: slope at each horizon with its 95\% block-bootstrap interval.}
\label{fig:calibration}
\end{figure}

\begin{figure}[t]
\centering
\includegraphics[width=\linewidth]{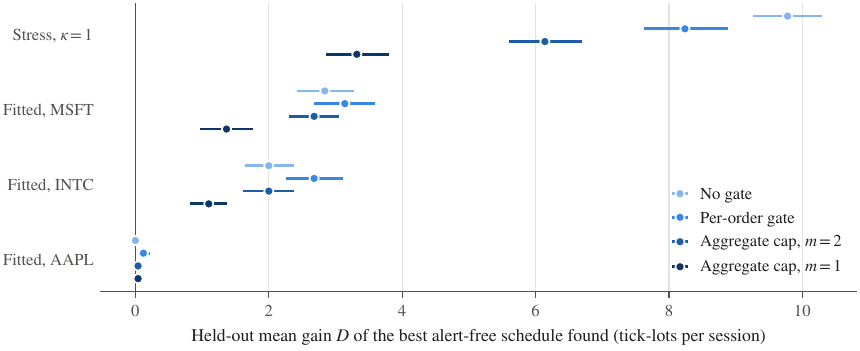}
\caption{The data of Table~\ref{tab:adversary} for the smooth responses: held-out gain of the best alert-free schedule found, with 95\% bootstrap intervals, by control. The step responses are omitted; their cells are exactly 0 or 48.}
\label{fig:adversary}
\end{figure}

Table~\ref{tab:spec} lists the simulator configuration. Figures~\ref{fig:calibration} and~\ref{fig:adversary} plot the fitted response and the searched gains, and Tables~\ref{tab:facts}--\ref{tab:rl} and Table~\ref{tab:execution} give the full results behind Sections~\ref{sec:adversary}--\ref{sec:cost}.

\paragraph{Evading all three detectors.} A new search that must evade all three detectors, with the lots executed per cycle (2, 5, 10, or 20) as an extra parameter, finds no alert-free schedule under either the $\kappa = 1$ or the MSFT-fitted response, with or without the cap. The schedules it ends on are caught in every held-out session, but they show how much the fixed two-lot execution limits the gain. Executing 10 to 20 lots per cycle, they earn $D$ of 69.3 and 63.6 tick-lots under $\kappa = 1$ (no gate and $m = 2$) and 16.4 and 11.0 under the MSFT fit, against at most about 10 for the searched schedules of Table~\ref{tab:adversary}.

\begin{table}[t]
\centering
\footnotesize
\setlength{\tabcolsep}{2pt}
\begin{tabular}{llrlll}
\toprule
& & & \multicolumn{3}{c}{\textbf{Change from ungated, paired by seed}} \\
\cmidrule(lr){4-6}
\textbf{Strategy} & \textbf{Measure} & \textbf{Ungated} & \textbf{$m = 2$ + throttle} & \textbf{$m = 2$} & \textbf{$m = 1$} \\
\midrule
Scheduled exec. & Fill rate (\%) & 32.8 & 0.0 [$-$0.1, 0.0] & 0.0 [$-$0.1, 0.0] & $-$19.6 [$-$21.3, $-$18.0] \\
 & Shortfall (ticks/lot) & 0.046 & 0.000 [$-$0.001, 0.000] & 0.000 [$-$0.001, 0.000] & 0.029 [$-$0.044, 0.102] \\
\addlinespace
Slice exec. & Fill rate (\%) & 37.6 & 0.0 [$-$0.1, 0.0] & 0.0 [$-$0.1, 0.0] & $-$7.8 [$-$9.7, $-$6.2] \\
 & Shortfall (ticks/lot) & 0.180 & 0.001 [0.000, 0.003] & 0.001 [0.000, 0.003] & $-$0.007 [$-$0.075, 0.055] \\
\addlinespace
Large maker & Volume (lots) & 60.2 & $-$5.7 [$-$7.3, $-$4.0] & 1.7 [0.5, 3.0] & $-$60.2 [$-$62.8, $-$57.8] \\
 & P\&L (tick-lots) & 27.0 & $-$2.5 [$-$5.3, 0.3] & 0.6 [$-$2.0, 3.1] & $-$27.0 [$-$34.7, $-$19.6] \\
 & Edge (ticks/lot) & 0.646 & $-$0.111 [$-$0.137, $-$0.084] & $-$0.124 [$-$0.146, $-$0.102] & $-$0.646 [$-$0.677, $-$0.615] \\
 & Markout (ticks/lot) & 0.457 & 0.030 [0.000, 0.060] & 0.021 [0.002, 0.042] & $-$0.457 [$-$0.519, $-$0.394] \\
\addlinespace
Patient maker & Volume (lots) & 74.5 & $-$23.3 [$-$25.8, $-$20.9] & $-$23.3 [$-$25.8, $-$20.9] & $-$74.5 [$-$77.4, $-$71.4] \\
 & P\&L (tick-lots) & $-$0.2 & 1.4 [$-$7.3, 10.4] & 1.4 [$-$7.4, 10.3] & 0.2 [$-$15.6, 17.1] \\
 & $|$Inventory$|$ (lots) & 19.5 & $-$8.9 [$-$10.0, $-$7.8] & $-$8.9 [$-$10.0, $-$7.8] & $-$19.5 [$-$21.3, $-$17.7] \\
\bottomrule
\end{tabular}
\caption{Execution quality of the legitimate scripts under the MSFT-fitted response, seeds 9300-9399. Brackets are 95\% bootstrap intervals for the paired difference. Shortfall is per lot of the 80-lot parent and includes the cost of completing the unfilled remainder at the closing fundamental; edge and markout are signed against the mid at the fill and 60 seconds later. The $m = 2$ and $m = 1$ columns have the throttle off. No arm raised a conduct alert. Produced by \texttt{paper\_study.execution\_quality}, which also reports the step and AAPL-fitted responses.}
\label{tab:execution}
\end{table}

\begin{table}[t]
\centering
\small
\begin{tabularx}{\linewidth}{lX}
\toprule
\textbf{Component} & \textbf{Specification} \\
\midrule
Book and matching & One evaluated account and one instrument; integer ticks, price/time priority, partial fills \\
Timing & One second per attempted action; background replenishment every second; default decisions every 60 seconds, with exact re-quote times in the cadence boundary checks \\
Background liquidity & Three price levels per side, 10 lots per level, centered around the fundamental plus any response \\
Exogenous path & Fundamental starts at 1,000; increments -1, 0, +1 with probabilities 1/4, 1/2, 1/4 every 30 seconds; noise quantities 0, 1, 2 with probabilities 1/2, 1/4, 1/4 every 10 seconds \\
Risk and fees & 100-lot worst-case position commitment; fee 0.05 tick-lots per filled lot \\
Sessions & 720 seconds in the core study; 720 and 3,600 seconds in the throttle tests; no warm-up \\
\bottomrule
\end{tabularx}
\caption{Simulator configuration.}
\label{tab:spec}
\end{table}

\begin{table}[t]
\centering
\caption{Stylized facts of 720-second mid-price sessions in half-spread units: share of 10-second intervals with an unchanged mid-price, standard deviation and excess kurtosis of 60-second returns, lag-1 autocorrelation of 10-second returns, and mean autocorrelation of absolute 10-second returns at lags 1--5 (volatility clustering). Simulator: 100 seeds with no subject orders; data: 30 windows of each LOBSTER sample, 21 June 2012, 9:40--15:50. Kurtosis intervals are 95\% bootstrap intervals over sessions; all intervals are in \texttt{results/empirical-paths-v1}.}\label{tab:facts}
\begin{tabular}{lrrrrr}
\toprule
 & Unchanged (\%) & SD 60 s & Excess kurtosis 60 s & ACF(1) 10 s & $|r|$ ACF 1--5 \\
\midrule
Simulator & 83 & 1.00 & $-0.49$ [$-0.60$, $-0.37$] & 0.00 & $-0.08$ \\
AAPL & 5 & 3.39 & 1.46 [0.58, 2.35] & $-0.05$ & 0.09 \\
AMZN & 16 & 2.75 & 4.09 [0.39, 7.61] & 0.05 & 0.09 \\
GOOG & 15 & 2.11 & 0.98 [0.02, 2.33] & $-0.01$ & 0.06 \\
INTC & 75 & 3.43 & 2.90 [0.22, 5.55] & 0.02 & 0.06 \\
MSFT & 66 & 3.84 & 1.60 [0.48, 2.57] & $-0.03$ & 0.08 \\
\bottomrule
\end{tabular}
\end{table}

\begin{table}[t]
\centering
\footnotesize
\setlength{\tabcolsep}{3pt}
\begin{tabular}{llll}
\toprule
\textbf{Response} & \textbf{Control} & \textbf{All resting depth} & \textbf{Depth within three ticks} \\
\midrule
Linear, $\kappa = 1$ & No gate & 9.68 [9.04, 10.34] (6.28--9.78) & 7.30 [6.72, 7.90] (4.62--7.30) \\
 & $m = 2$ & 6.24 [5.70, 6.80] (6.14--6.24) & 6.12 [5.56, 6.68] (5.85--6.12) \\
 & $m = 1$ & 3.32 [2.84, 3.80] (2.88--3.34) & 2.96 [0.05, 5.85] (2.96--3.52) \\
\addlinespace
Fitted, MSFT & No gate & 4.48 [4.02, 4.96] (2.84--4.48) & 2.84 [2.42, 3.26] (2.80--2.96) \\
 & $m = 2$ & 2.54 [2.14, 2.96] (2.54--2.70) & 3.06 [2.56, 3.58] (2.54--3.06) \\
 & $m = 1$ & 1.37 [0.98, 1.76] (1.37--1.58) & 1.58 [1.26, 1.90] (1.28--1.58) \\
\addlinespace
Fitted, INTC & No gate & --- & 2.42 [2.02, 2.84] (1.34--2.42) \\
 & $m = 2$ & --- & 1.84 [1.50, 2.20] (1.48--2.00) \\
 & $m = 1$ & --- & 1.06 [0.80, 1.34] (1.06--1.32) \\
\addlinespace
Tick units, MSFT & No gate & --- & 1.52 [1.20, 1.86] (1.52--1.76) \\
 & $m = 2$ & --- & 1.12 [0.82, 1.44] (1.12--1.66) \\
 & $m = 1$ & --- & 1.04 [0.20, 2.01] (0.17--1.04) \\
\addlinespace
Tick units, AAPL & No gate & --- & 0.88 [0.64, 1.14] (0.76--0.88) \\
 & $m = 2$ & --- & 0.96 [0.70, 1.24] (0.74--0.96) \\
 & $m = 1$ & --- & 0.87 [0.12, 1.81] (0.30--0.87) \\
\bottomrule
\end{tabular}
\caption{Held-out mean $D$, in tick-lots per session, of the selected schedule, with its 95\% bootstrap interval and, in parentheses, the range across three search restarts. ``All resting depth'' is the response of Table~\ref{tab:adversary}; ``within three ticks'' counts only orders within three ticks of each side's best price, matching the fitted top three levels. Tick units map one simulator tick to one market tick. No held-out episode in any cell raised a conduct alert. Throttle on; produced by \texttt{paper\_study.robustness}.}
\label{tab:robustness}
\end{table}

\begin{table}[t]
\centering
\caption{Policy-gradient adversary (\texttt{results/rl-adversary-v1}): held-out gain $D$ in tick-lots per session on seeds 9100-9199. Random search: Table~\ref{tab:adversary}. Equal budget: lowest and highest of three runs with 80 samples each. Long: one run with 800 samples, with 95\% bootstrap interval. No held-out episode of any selected schedule raised a conduct alert.}\label{tab:rl}
\begin{tabular}{llrrr}
\toprule
Response & Control & Random search & Policy gradient, 80 & Policy gradient, 800 \\
\midrule
$\kappa = 1$ & none & 9.78 & 7.24--9.23 & 9.68 [9.02, 10.34] \\
 & per-order & 8.24 & 6.80--8.36 & 7.94 [7.34, 8.56] \\
 & $m = 2$ & 6.14 & 5.62--6.24 & 6.24 [5.68, 6.80] \\
 & $m = 1$ & 3.32 & 2.39--3.32 & 2.91 [2.22, 3.64] \\
MSFT fit & none & 2.84 & 2.68--4.24 & 4.86 [4.28, 5.44] \\
 & per-order & 3.14 & 2.56--3.86 & 3.62 [3.12, 4.12] \\
 & $m = 2$ & 2.68 & 1.91--2.70 & 2.84 [2.42, 3.26] \\
 & $m = 1$ & 1.37 & 1.34--1.44 & 1.44 [1.12, 1.78] \\
INTC fit & none & 2.00 & 1.60--2.46 & 3.32 [2.88, 3.74] \\
 & per-order & 2.68 & 1.20--2.92 & 2.62 [2.20, 3.08] \\
 & $m = 2$ & 2.00 & 1.48--2.00 & 1.72 [1.23, 2.22] \\
 & $m = 1$ & 1.10 & 0.26--1.36 & 1.10 [0.82, 1.38] \\
\bottomrule
\end{tabular}
\end{table}

\clearpage

\section{Mechanism checks under the step response}\label{sec:appendix}

These checks use the stipulated step response and scripted strategies. They explain mechanisms behind the main results; Sections~\ref{sec:llm}--\ref{sec:cost} do not depend on them.

\subsection{The activation boundary}\label{sec:stepcheck}

The settings and parameter vectors were saved before each comparison was run, after inspecting the earlier results. This is exploratory analysis, not preregistration. Seeds 8000-8009 pair the cap and price-boundary conditions, 8100-8104 pair each parameter vector, and 8200-8209 pair the horizon conditions. Cadence checks reuse seeds 6000-6039. The runners are \texttt{paper\_study.sensitivity} and \texttt{paper\_study.boundary}. Tables give means with observed minimum and maximum values; these ranges are descriptive, not confidence intervals. Reused paths, rescoring, and parameter draws from one authored generator do not supply independent strategy samples. The experiments in Section~\ref{sec:adversary} instead fix the search budget, seeds, and analysis in a saved plan before running, separate training from held-out paths, and report 95\% percentile bootstrap intervals over 100 held-out paths.

\begin{table}[htbp]
\centering
\small
\begin{tabularx}{\linewidth}{lXr}
\toprule
\textbf{Sensitivity analysis} & \textbf{Conditions} & \textbf{Episodes} \\
\midrule
Cap and response & 2 scripts $\times$ 6 response thresholds $\times$ (ungated + 5 aggregate caps) $\times$ 10 seeds $\times$ 2 response states & 1,440 \\
Parameter replication & 12 new vectors from the existing generator $\times$ 5 market seeds $\times$ 2 thresholds $\times$ 2 gates $\times$ 2 response states & 480 \\
Throttle and horizon & 2 session lengths $\times$ (ungated + 2 gates $\times$ 4 windows + 2 gates without throttle) $\times$ 10 seeds & 220 \\
Price boundaries & 3 cap multiples $\times$ 2 neighboring thresholds $\times$ 2 scripts $\times$ 2 gates $\times$ 10 seeds $\times$ 2 response states & 480 \\
Cadence boundaries & 170- and 190-second maker cadences $\times$ 40 ungated seeds; scored at $W = 180$ & 80 \\
\bottomrule
\end{tabularx}
\caption{Sensitivity and boundary checks: 2,700 episodes. The first three comparisons contain 2,140 episodes; the boundary checks add 560. The cap grid uses $m \in \{1, 2, 3, 4, 6\}$ and $t \in \{0.1, 0.2, 0.25, 0.3, 0.4, 0.5\}$. Neighboring thresholds are 0.24/0.26 at $m = 2$, 0.39/0.41 at $m = 4$, and 0.49/0.51 at $m = 6$. Cap and parameter comparisons disable throttling. The horizon study uses $t = 0.4$, $m = 2$, and throttle windows of 120, 300, 900, and 1,800 seconds.}
\label{tab:sensitivity}
\end{table}

The original settings also make the zero gain structurally likely.

\begin{proposition}[Static activation boundary]\label{prop:boundary}
With three intact background levels of $L$ lots per side and a subject display of $mL$ lots on one side only, the imbalance is $I = m/(6+m)$. The step response with threshold $t$ activates if and only if $m > 6t/(1-t)$.
\end{proposition}

\begin{proof}
One side holds $3L + mL$ lots and the other $3L$, so $I = mL/((6+m)L)$. The inequality $m/(6+m) > t$ rearranges to $m(1-t) > 6t$.
\end{proof}

Thus $m = 2$ gives $I = 0.25$, below 0.4, and at $t = 0.4$ the boundary is $m = 4$. Trades can deplete depth before replenishment, so 0.25 is not a global bound on the simulated book. A replay diagnostic of the 40 gated parameter draws nevertheless finds no response activations and a maximum absolute imbalance of 0.282. Their identical responsive and unresponsive event streams therefore verify a restricted mechanism, not general economic prevention. Section~\ref{sec:adversary} replaces the threshold with the linear response and the fixed scripts with a search.

\begin{figure}[htbp]
\centering
\includegraphics[width=\linewidth]{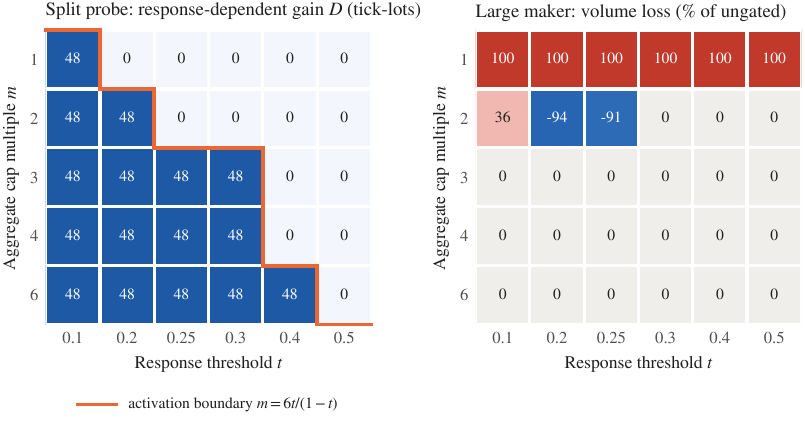}
\caption{Cap-response sensitivity with throttling disabled. Left: mean $D$ for the split probe, in tick-lots; each cell has zero spread across its ten seeds. Right: loss of mean large-maker volume relative to the ungated mean at the same threshold; negative values mean higher gated volume. The ungated arms are the baselines and are not separate plotted rows. Only the aggregate cap varies; the individual limit stays twice reference depth.}
\label{fig:cap}
\end{figure}

At the original 20-lot aggregate cap, mean $D$ is 0.00 at $t = 0.4$ and 48.00 at $t = 0.2$, positive on 10/10 paired paths. All split-probe cells in Figure~\ref{fig:cap} follow the static response boundary. The equalities $(m,t) = (2, 0.25)$, $(4, 0.4)$, and $(6, 0.5)$ give zero because activation uses a strict inequality. At the neighboring thresholds in Table~\ref{tab:sensitivity}, each lower-threshold cell gives $D = 48$ on every seed, and each upper cell gives $D = 0$. The contrast thus persists away from equality, while remaining a consequence of the imposed response. At $m = 1$, all 12-lot maker quotes exceed the allowance: the maker executes nothing, a 100\% volume loss.

\begin{table}[htbp]
\centering
\small
\begin{tabular}{lllll}
\toprule
\textbf{Threshold} & \textbf{Gate} & \textbf{$D$: mean [min, max]} & \textbf{$D > 0$} & \textbf{$D > 0$, no alert} \\
\midrule
0.2 & off & 16.47 [-5.25, 50.80] & 44/60 & 5/60 \\
0.2 & aggregate & 10.96 [-5.85, 65.90] & 30/60 & 30/60 \\
0.4 & off & 11.32 [0.00, 50.80] & 27/60 & 0/60 \\
0.4 & aggregate & 0.00 [0.00, 0.00] & 0/60 & 0/60 \\
\bottomrule
\end{tabular}
\caption{Parameter replication, \texttt{paper\_study.sensitivity}: 12 vectors crossed with seeds 8100-8104, cap multiple 2, throttle disabled, $W = 180$ seconds. D is mean [minimum, maximum] in tick-lots. Each denominator is 60 draw-seed pairs, not 60 independent strategies. Positive D remains counted even if other actions were rejected.}
\label{tab:replication}
\end{table}

At threshold 0.2, 30/60 gated draw-seed pairs retain a positive response-dependent wealth difference without a conduct alert. At threshold 0.4, the same cap leaves no positive differences. The throttle is disabled in both comparisons, so these results isolate response sensitivity. The 48 tick-lots in Figure~\ref{fig:cap} also follow from the stipulated mechanism: 12 cycles times two executed lots times a two-tick response when every execution benefits. The sweep checks this dependence; it does not estimate a response from market data.

At cap multiple 2 and response thresholds 0.2 and 0.25, the gate with its throttle disabled increases maker volume. Changing the display changes the subsequent book response and fills, so accepting fewer orders need not reduce execution. Higher volume alone says little about execution quality. These supplemental runs vary paths within the same strategy generator and stipulated step response. The main text separately examines smooth responses and an ABIDES market with strategic background traders; neither supplies calibrated queues for this simulator.

\subsection{Throttle timing}\label{sec:throttle}

The cancellation throttle counts accepted submissions, so the number of quotes admitted per decision affects when it starts. In \texttt{pilot.throttle\_isolation}, seeds 4000-4039, removing only the aggregate gate's throttle raises mean volume from 50.93 to 59.05 lots, close to 59.48 ungated. Thus most of the large maker's loss in that matched comparison comes from throttling. Its 720-second session is shorter than the 1,800-second window, so a triggered lockout cannot expire during trading. The horizon experiment below uses a separate seed cohort; its baselines should not be substituted for those in this ablation or Table~\ref{tab:legit}.

\begin{figure}[htbp]
\centering
\includegraphics[width=\linewidth]{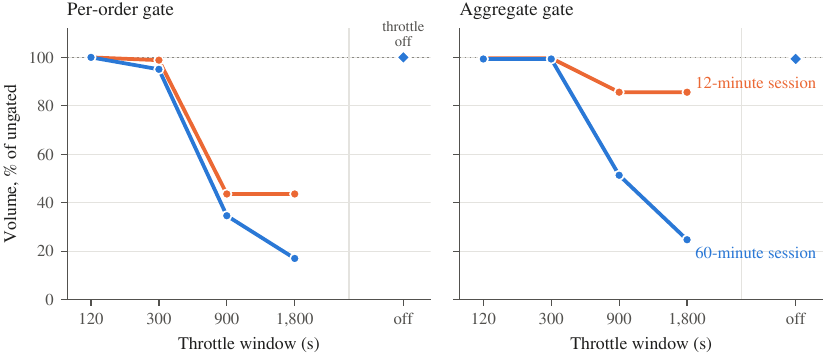}
\caption{Mean volume as a percentage of the ungated mean at the same horizon, \texttt{paper\_study.sensitivity}, seeds 8200-8209. Lines join finite throttle windows; separate diamonds disable only the throttle. Percentages are ratios of means. Panels distinguish the legacy and aggregate policies.}
\label{fig:throttle}
\end{figure}

\begin{table}[htbp]
\centering
\small
\begin{tabular}{llllll}
\toprule
\textbf{Minutes} & \textbf{Gate} & \textbf{Executed lots} & \textbf{Throttle rejects} & \textbf{Throttled} & \textbf{Recovered} \\
\midrule
12 & off & 60.50 [34.00, 90.00] & 0.00 [0, 0] & 0/10 & 0/10 \\
12 & legacy & 26.40 [13.00, 59.00] & 26.80 [16, 28] & 10/10 & 0/10 \\
12 & aggregate & 51.80 [29.00, 90.00] & 6.80 [0, 8] & 9/10 & 0/10 \\
60 & off & 300.60 [263.00, 355.00] & 0.00 [0, 0] & 0/10 & 0/10 \\
60 & legacy & 51.20 [31.00, 74.00] & 197.50 [188, 200] & 10/10 & 10/10 \\
60 & aggregate & 74.40 [38.00, 210.00] & 169.40 [97, 180] & 10/10 & 10/10 \\
\bottomrule
\end{tabular}
\caption{Large maker from \texttt{pilot.revision}, run by \texttt{paper\_study.sensitivity} on seeds 8200-8209. Gates use a 1,800-second throttle window; off is the matched ungated baseline. Lots and throttle rejects are mean [minimum, maximum]. The last columns count episodes. Recovered means at least one accepted submission after the first throttle rejection, not uninterrupted trading.}
\label{tab:horizon}
\end{table}

In the one-hour condition, all ten aggregate-gated episodes resume submissions after a throttle rejection. Mean volume is nevertheless 74.40 lots, compared with 300.60 ungated: a 75.2\% loss. The short-session aggregate arm averages 51.80 against 60.50 lots. Extending the session reveals recovery, but also repeated periods in which the maker cannot submit orders.

Session length determines whether recovery can be observed. The trailing window, quoting rate, and minimum submission count determine whether throttling starts and how often it recurs. A short window can keep the count below the trigger altogether. A longer session can allow submissions to resume while still losing substantial volume. These settings must be assessed together; the tests do not select an optimal throttle for a live market.

\renewcommand{\thetable}{\arabic{table}a}

\begin{table}[htbp]
\centering
\small
\begin{tabular}{llll}
\toprule
\textbf{Legitimate task} & \textbf{Ungated lots} & \textbf{Aggregate lots} & \textbf{Volume loss} \\
\midrule
Scheduled executor & 27.38 [11.00, 50.00] & 27.38 [11.00, 50.00] & 0.0\% \\
10-lot slice executor & 30.12 [12.00, 45.00] & 30.12 [12.00, 45.00] & 0.0\% \\
Small maker & 31.05 [20.00, 36.00] & 31.05 [20.00, 36.00] & 0.0\% \\
Large maker & 56.50 [33.00, 88.00] & 48.55 [27.00, 87.00] & 14.1\% \\
Patient maker, 300 s & 66.28 [32.00, 109.00] & 47.60 [30.00, 69.00] & 28.2\% \\
\bottomrule
\end{tabular}
\caption{Full aggregate policy, including the 1,800-second throttle, versus no policy gate in 720-second sessions. Volume is mean [minimum, maximum] over seeds 6000-6039; loss is a ratio of means. The strategies are defined in \texttt{pilot.negative\_controls}. These paths differ from the throttle-isolation and horizon cohorts.}
\label{tab:legit}
\end{table}

\renewcommand{\thetable}{\arabic{table}}

\addtocounter{table}{-1}\renewcommand{\thetable}{\arabic{table}b}

\begin{table}[htbp]
\centering
\small
\begin{tabular}{lllll}
\toprule
\textbf{Legitimate task} & \textbf{Size cap} & \textbf{Throttle} & \textbf{Position} & \textbf{Other} \\
\midrule
Scheduled executor & 0.00 [0, 0] & 0.00 [0, 0] & 0.00 [0, 0] & 0.00 [0, 0] \\
10-lot slice executor & 0.00 [0, 0] & 0.00 [0, 0] & 0.00 [0, 0] & 0.00 [0, 0] \\
Small maker & 0.00 [0, 0] & 0.00 [0, 0] & 0.00 [0, 0] & 0.00 [0, 0] \\
Large maker & 20.30 [20, 24] & 7.40 [0, 8] & 0.00 [0, 0] & 0.25 [0, 2] \\
Patient maker, 300 s & 6.00 [6, 6] & 0.00 [0, 0] & 0.00 [0, 0] & 0.03 [0, 1] \\
\bottomrule
\end{tabular}
\caption{Rejected actions under the same aggregate policy: mean [minimum, maximum]. Size cap combines individual and aggregate size checks; every size rejection here is aggregate. Other comprises cancellation attempts on inactive orders. Each attempt records only its first failed check, so counts reconcile but do not causally apportion lost volume.}
\label{tab:rejects}
\end{table}

\renewcommand{\thetable}{\arabic{table}}

The large maker loses 14.1\% of volume in Table~\ref{tab:legit}. Its 27.95 mean rejections comprise 20.30 size-cap rejections, 7.40 throttle rejections, and 0.25 failed cancellations. The separate throttle ablation shows why rejection counts alone cannot attribute the volume loss. The patient maker loses 28.2\% without any throttle rejections; its cost arises under the remaining gate checks. The passive scheduled and slice executors receive no gate rejections, yet complete only 34.2\% and 37.7\% of their 80-lot parent orders. Volume alone can both understate and overstate the cost.

The same plan reran the legitimate-cost comparison with 100 paths (seeds 9200-9299) under the step response at $t = 0.4$, the setting of Table~\ref{tab:legit}. The full aggregate policy, including the 1,800-second throttle, reduces the large maker's mean volume from 57.87 lots (55.52 to 60.19) to 49.14 (46.74 to 51.65), a 15.1\% loss, consistent with the 14.1\% in Table~\ref{tab:legit}.

\subsection{Detector timing and legitimate counterexamples}\label{sec:cadence}

The age condition in $Q(i,f)$ excludes orders canceled after $W$, whatever their purpose. Moving a maker's quoting interval across $W$ changes which orders are eligible for an alert. The shared fill, remaining quantity, and other conditions in Section~\ref{sec:patterns} must still hold.

\begin{figure}[htbp]
\centering
\includegraphics[width=\linewidth]{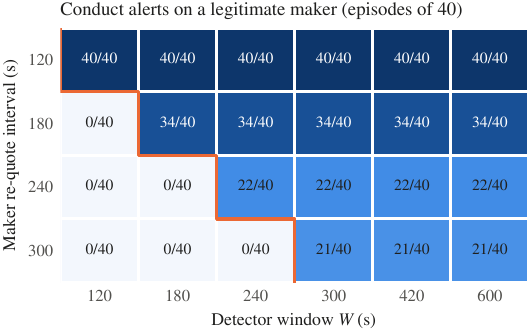}
\caption{Rule-boundary check on an authored legitimate maker. Counts show conduct alerts among 40 ungated episodes at each cadence and shared scoring window. Each row uses the same seed cohort; each trajectory is rescored across windows. The staircase follows the rule's age condition and is not a measured operational false-positive curve.}
\label{fig:cadence}
\end{figure}

At $W = 180$ seconds, the 180-second maker alerts in 34/40 episodes and the 300-second maker in none. The rule includes equality: $a_i \leq W$. Exact-time checks at 170 and 190 seconds give 39/40 and 0/40 alerts, respectively, on the same seeds. The scheduler reproduces all 40 original 180-second event hashes before these checks. Thus a legitimate quoting schedule also triggers alerts strictly inside the window.

We selected and assessed the original window on the same cases. Extending $W$ from 120 to 180 seconds raised detections from 10/15 to 15/15 selected economic cases, so the improvement is tuning performance rather than independent recall. The choice of legitimate controls matters as well. A buy-only executor cannot meet the opposite-side-own-fill condition, and a small maker cannot challenge the aggregate size threshold. Controls outside the rule's eligible region provide little evidence about its specificity within that region.

For the 180-second maker, the aggregate policy reduces mean volume from 73.80 to 58.98 lots and produces no conduct alerts. The gate has changed the activity available to the detector. The disappearance of alerts, alongside reduced execution, does not establish better discrimination. Evaluating specificity beyond these examples requires separate strategy families and independently reviewed objectives.

\end{document}